\documentclass[num-refs]{wiley-article}

\usepackage{siunitx}
\usepackage{multicol}
\graphicspath{{./Figures/}}

\papertype{Original Article}

\title{Assessment of protein assembly prediction in CASP13}

\author[1]{Dmytro Guzenko}
\author[2]{Aleix Lafita}
\author[3]{Bohdan Monastyrskyy}
\author[2]{Andriy Kryshtafovych}
\author[1]{Jose M. Duarte}

\affil[1]{Research Collaboratory for Structural Bioinformatics Protein Data Bank, San Diego Supercomputer Center, University of California, San Diego, La Jolla, CA 92093, USA}
\affil[2]{European Molecular Biology Laboratory, European Bioinformatics Institute, Wellcome Genome Campus, Hinxton, Cambridge, CB10 1SD, UK}
\affil[2]{Protein Structure Prediction Center, Genome and Biomedical Sciences Facilities, University of California, Davis, CA 95616, USA}

\corraddress{Dmytro Guzenko, Research Collaboratory for Structural Bioinformatics Protein Data Bank, San Diego Supercomputer Center, University of California, San Diego, La Jolla, CA 92093, USA}
\corremail{dmytro.guzenko@rcsb.org}

\fundinginfo{DG and JD were supported by the RCSB PDB, jointly funded by the National Science Foundation, the National Institute of General Medical Sciences, the National Cancer Institute, and the Department of Energy (NSF-DBI 1338415; Principal Investigator: Stephen K. Burley). AK and BM were supported by the US National Institute of General Medical Sciences (NIGMS/NIH) grant GM100482. AL was supported by the EMBL International PhD Programme.}

\runningauthor{Guzenko et al.}

\begin{document}

\maketitle

\begin{abstract}
We present the assembly category assessment in the 13\textsuperscript{th} edition of the CASP community-wide experiment. For the second time, protein assemblies constitute an independent assessment category. Compared to the last edition we see a clear uptake in participation, more oligomeric targets released, and consistent, albeit modest, improvement of the predictions quality. Looking at the tertiary structure predictions we observe that ignoring the oligomeric state of the targets hinders modelling success. We also note that some contact prediction groups successfully predicted homomeric interfacial contacts, though it appears that these predictions were not used for assembly modelling. Homology modelling with sizeable human intervention appears to form the basis of the assembly prediction techniques in this round of CASP. Future developments should see more integrated approaches to modelling where multiple subunits are a natural part of the modelling process, which would benefit the structure prediction field as a whole.

\end{abstract}



\begin{multicols}{2}
\section{Introduction}

In their physiological environment, protein chains commonly associate with other chains or copies of themselves to form protein assemblies. This is the so-called \emph{quaternary} structure, an intrinsic property of the native state of a protein, known before the first atomic structures were solved \cite{svedberg1927application}.
Protein function is linked and often is determined or regulated by the oligomeric structure \cite{goodsell2000structural}\cite{selwood2012dynamic}\cite{hashimoto2010mechanisms}.
As of March 2019, the average structure in the Protein Data Bank (PDB) \cite{burley2018rcsb} is a dimer and approximately half of the PDB is annotated as oligomeric.
Estimates of the average protein oligomeric state in the cell point to an even higher tetrameric assembly \cite{goodsell1991inside}.

Protein oligomerization is a broad term that encompasses states with different degrees of affinity. The association between polypeptide chains in stable obligate oligomers can be regarded as an extension of protein folding and often occurs simultaneously \cite{tsai1998protein}. At the other extreme are transient protein-protein complexes where the association is opportunistic and promiscuous, representing the functions of the proteins involved \cite{nooren2003diversity}. It is important to note that there is a continuum between these states, and in the context of CASP no effort has yet been made to distinguish them.

Due to intrinsic limitations of the different experimental methods used for structure determination, protein assemblies are likely underrepresented in the PDB. The three methods most commonly used are X-ray crystallography, nuclear magnetic resonance (NMR) spectroscopy and 3-dimensional electron microscopy (3DEM). 

X-ray crystallography has been and remains the main source of atomic-resolution protein structures in the PDB. The majority of these are homomeric (85\% of depositions in 2018), from which about half are oligomeric. Crystallization of hetero-oligomers is more technically challenging, especially as the interaction becomes more transient \cite{radaev2006survey}. Consequently, hetero-oligomeric complexes are severely underrepresented in the X-Ray crystallographic output. 

Historically the second-most popular method for protein structure determination, NMR spectroscopy, does not contribute significantly to their oligomerization knowledge. It accounted for 3\% of overall depositions to PDB in 2018 with 90\% of entries being monomers. The reasons are mostly technical: protein complexes are often large and symmetric and both of these factors complicate NMR data analysis.

The rapidly expanding 3DEM technique is naturally suited for determination of protein complexes (95\% of the EM entries) and has the most potential to boost our quaternary structure knowledge. In 2018 3DEM accounted for 10\% of PDB depositions and, notably, for about a third of all deposited hetero-oligomeric complexes. 
Traditionally, the interpretation of the experimental maps was more challenging due to low resolution (median 4.3 $\r{A}$) and less well-developed data-model fit quality metrics. However there is plenty of room for optimism as the technique continues to actively develop and achieves ever higher resolutions (the median resolution was 3.8 $\r{A}$ in 2018) \cite{kuhlbrandt2014resolution}\cite{ognjenovic2019frontiers}.

The Critical Assessment of protein Structure Prediction (CASP) experiment was established as a means to consistently evaluate the state of the protein structure computational modeling field. 
The experiment focuses on problems at the frontier of the research and evolves together with it. New prediction categories deemed attainable are regularly introduced, and those where the progress is believed to have been exhausted are discontinued \cite{kryshtafovych2010casp}. 

Quaternary structure has a rather peculiar history within the experiment. While oligomeric protein targets were incidentally featured in CASP2 (1996), CASP7 (2006) and CASP9 (2010), the experiment was mainly focused on tertiary structure prediction. On the other hand, the Critical Assessment of PRedicted Interactions (CAPRI), an independent experiment inspired by CASP, was established in 2001 to address the protein-protein docking problem. With such an arrangement, the assessment  of the quaternary structure modeling was explicitly branched into “subunits” (CASP) and “interfaces” (CAPRI). Recognizing the growing importance of integrated quaternary structure prediction, CASP and CAPRI conducted the parallel assessment of selected oligomeric targets in 2014 (CASP11/CAPRI30). In 2016 (CASP12), a separate “Assembly” category was introduced to evaluate predictions of the complete 3-dimensional functional units on all oligomeric CASP targets. The assembly category serves to highlight the importance of considering proteins in their native solution state, with the ultimate goal of producing complete models, that can shed light into the biology and function of the molecular systems under scrutiny.

By introducing new assessment categories, the CASP experiment shapes and drives the development of methods necessary to excel in them \cite{kryshtafovych2010casp}. Recent breakthroughs in both domain structure \cite{abriata2018assessment} and contact predictions \cite{schaarschmidt2018assessment} suggest that higher-order complexity targets, protein assemblies, are feasible. Here we present our analysis of the CASP13 assembly predictions, compare the results to those of CASP12 and discuss the status and outlook of the field.


\section{Methods}

\subsection{Assembly targets}
In CASP13, the organizers proactively gathered protein assemblies, specifically targeting heteromeric complexes. This has resulted in 64\% of the targets (42 out of 66) being oligomeric -- a marked increase from 42\% (30 targets out of 71) in CASP12\cite{moult2018critical}. 20 targets were selected for the combined CAPRI/CASP experiment [ref CAPRI assessment].

In terms of experimental methods the vast majority of targets came from X-ray crystallography (36 out of 42), whilst the rest were solved with the 3DEM technique. Compared to CASP12 (26 X-ray, 2 NMR and 2 3DEM) we observe a significant increase in structures solved with 3DEM, consistent with the recent developments in experimental structural biology.

Assigning the oligomeric state of targets was not always a straightforward task, specifically in the case of crystal structures, where the contacts in the crystal lattice can lead to different interpretations \cite{capitani2015understanding}. This step was done in collaboration with the CAPRI assessment team, with contributions from the CASP organizers. In broad terms, to assign the oligomeric state we considered the following (in order of priority):
\begin{enumerate}
    \item experimentalists indication, preferred if backed by experimental evidence;
    \item if structure was known, EPPIC \cite{bliven2018automated} and PISA \cite{krissinel2015stock} analysis;
    \item stoichiometry consensus of homologous structures in the PDB found with HHpred \cite{zimmermann2018completely}.
\end{enumerate}
All CASP13 targets were examined in this way, even when assumed to be monomers by the experimentalists. After this procedure, 5 cases remained ambiguous and were assigned with low confidence (see Table S1). This shows how one of the challenges in assembly prediction is the definition of the ground truth \cite{capitani2015understanding}. 

The selection process resulted in a wide range of stoichiometries and symmetries (see Table S1). They included a helical symmetry (\textbf{T0995}) and a very large complex with A6B6C6 stoichiometry (\textbf{H1021}) solved by 3DEM. Out of 42 targets, 12 were heteromeric and 30 homomeric, double the proportion of heteromers as would be expected if drawn randomly from the PDB\cite{xu2019principles}. Two of the heteromeric targets presented uneven stoichiometry (\textbf{H0953}, with stoichiometry A3B1 and \textbf{H1022} with A6B3), a rather unusual event in the PDB with only 10\% occurrence among all known heteromers\cite{xu2019principles}.

\subsection{Target difficulty}
We have classified the targets into three difficulty levels based on the information available to the predictors prior to the experiment, similarly to the CASP12 assembly assessment\cite{lafita2018assessment}. Outcome of predictions (\emph{i.e.}, \emph{posterior} difficulty) was not considered.

We define three difficulty classes with the following criteria:
\begin{itemize}
    \item \textbf{Easy}: the target has templates for both the subunits and the overall assembly, findable by sequence homology detection methods.
    \item \textbf{Medium}: the target has partial templates identifiable by sequence homology detection methods. Partial can mean that the full subunit templates are known but no information to model the interface can be found, or that information of only part of the interfaces is known (e.g. a dimer template available for half of a tetrameric target).
    \item \textbf{Difficult}: the target does not have templates findable by sequence homology detection methods, for either the subunits or the assembly.
\end{itemize}

One of the targets (\textbf{T0965}) was classified as Medium (see Table S1), despite availability of a complete template, because the arrangement of helices at the interface differed substantially in the target structure.

\subsection{Evaluation scores}
\begin{figure}
\centering
\includegraphics[width=0.6\textwidth]{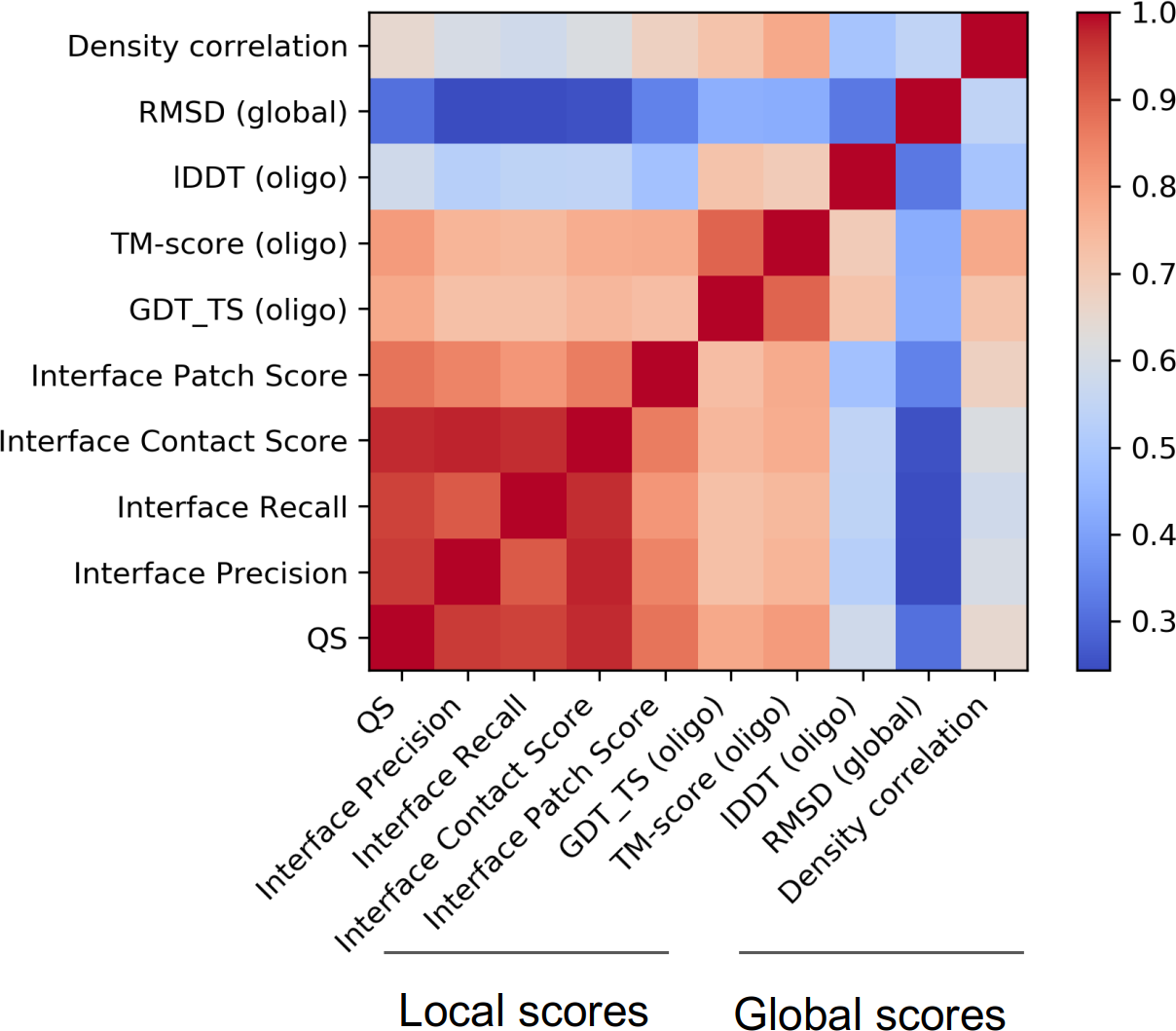}
\caption{Score correlations. A heat map with correlations among all relevant scores used in the predictioncenter.org web site. The “local” block of scores captures interface features, the “global” block captures features of the whole assembly.}
\label{figure:1:correlations}
\end{figure}

We assess the accuracy of the predicted protein-protein interfaces with the two measures introduced in the CASP12 assembly assessment: Interface Contact Similarity (ICS) and Interface Patch Similarity (IPS) \cite{lafita2018assessment}. In the official evaluation tables in the \url{predictioncenter.org} website, these scores are called F1 and Jaccard respectively. Evaluation of the interfaces is sufficient if the subunits are known or are relatively easy to model independently of each other. However, CASP assembly targets are not selected with this assumption in mind and in practice often require non-trivial subunit modelling. To capture performance of the tertiary structure prediction methods in the context of quaternary structure, we have chosen to add two other scores to the pool: local Distance Difference Test (lDDT) \cite{mariani2013lddt} for local model quality and Global Distance Test (GDT) \cite{zemla1999processing} for similarity of the global fold. These scores are not directly applicable to the multi-chain models, as the order of chains in the file is not necessarily preserved with respect to their 3-dimensional arrangement. Therefore, 'chain mapping' has to be established between the target and the prediction prior to regular scoring. We used the QS-score algorithm \cite{bertoni2017modeling} (all targets except \textbf{H1021}) and QS-align \cite{lafita2019biojava} (\textbf{H1021}) for this purpose. The obtained scores were rescaled to the $[0,1]$ range and are referred here as GDT/lDDT Oligomeric (or GDTo/lDDTo for brevity). In addition, we calculated these scores for the CASP12 targets and predictions to enable direct comparison of the results. Figure \ref{figure:1:correlations} shows score correlations for all models in CASP13, with clear blocks differentiating how interface (local) scores capture different information than assembly (global) scores.

\begin{figure}
\centering
\includegraphics[width=0.7\textwidth]{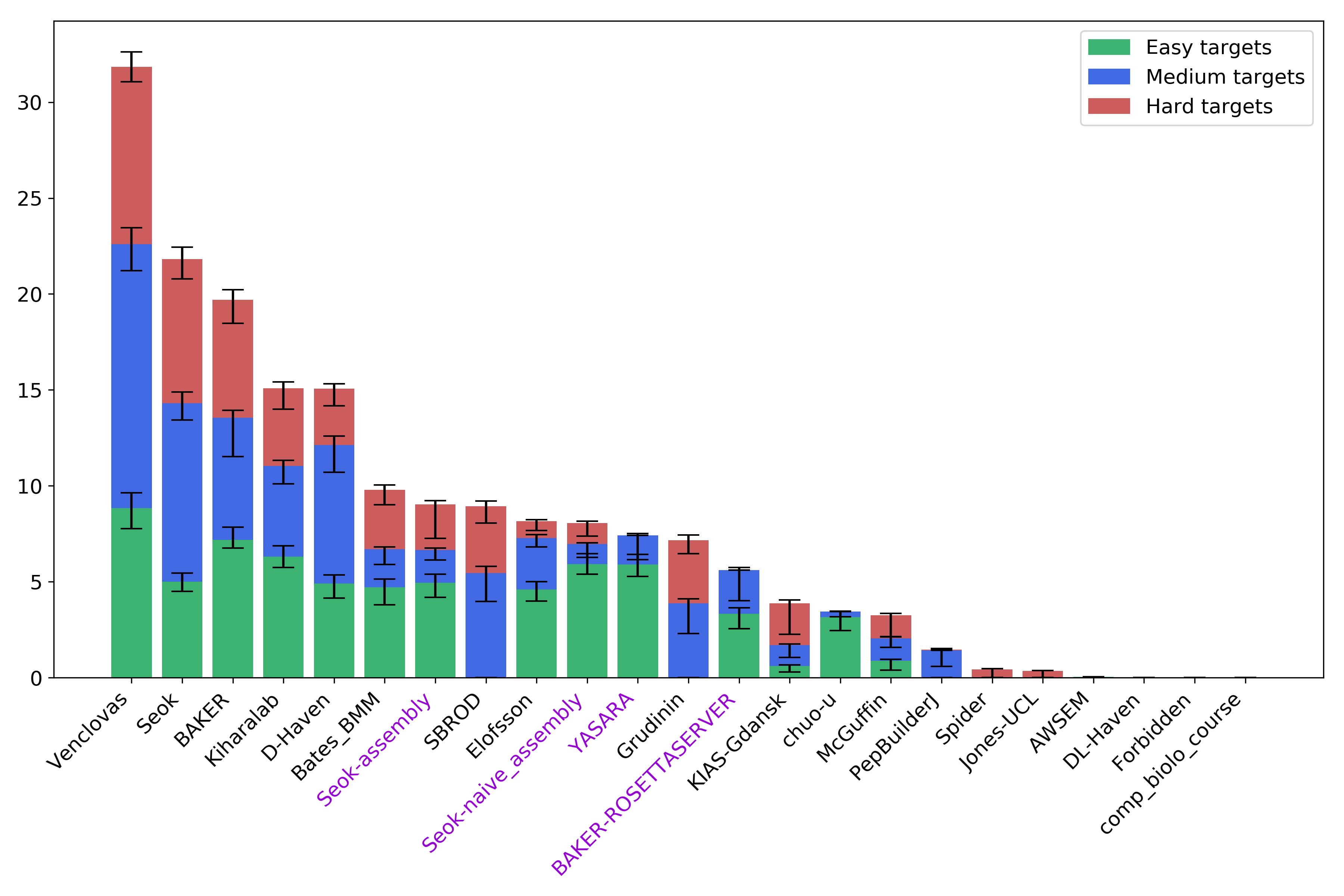}
\caption{Group rankings in the assembly category. The groups are sorted by the sum of $Z$-scores for all difficulty classes. The error bars are obtained by iteratively excluding every target from each difficulty class and recalculating the cumulative $Z$-scores. The server groups are labeled in violet.}
\label{figure:3:rankings}
\end{figure}

$Z$-scores were calculated for every score per evaluation target. The first submitted model (supposedly the best out of five allowed) was used for each group. To avoid penalizing unsuccessful prediction attempts and software glitches, we followed the CASP convention of removing outliers ($Z<-2$), recalculating the $Z$-scores and flattening negative values to zero. The total group score is a simple sum of all $Z$-scores for all targets it submitted predictions for. It has been noted \cite{cozzetto2009evaluation} that difficult targets with few good predictions may result in inflated $Z$-scores. To mitigate this effect we performed 'leave-one-out ranking', whereby each target is consecutively removed from consideration, and groups’ mean total score is used for the ranking. The maximum and minimum total score values can be used to assess the significance of the differences between the closely ranked groups (shown in Figure \ref{figure:3:rankings} as error bars).

\section{Results}

A total of 45 groups participated in the CASP13 assembly category. From those, 22 groups participated only in the subset of targets selected for the joint CASP/CAPRI experiment, while 23 submitted predictions for all targets. 17 groups submitted models for more than 10 targets. That compares to only 10 groups submitting models for more than 10 targets in CASP12 assembly category \cite{lafita2018assessment}. In terms of number of models submitted there was a dramatic increase from 1600 in CASP12 to more than 5000 in CASP13.

Clear improvements in the prediction format and methodology were introduced in this edition compared to the first assembly category experiment in CASP12. First, the stoichiometry information is now provided to the prediction servers in an automated way. Second, model files can now be multi-chain, eliminating the need for assessors to guess whether predictors are actually attempting assembly prediction or not.

\subsection{Performance}
\begin{figure}
\centering
\includegraphics[width=0.9\textwidth]{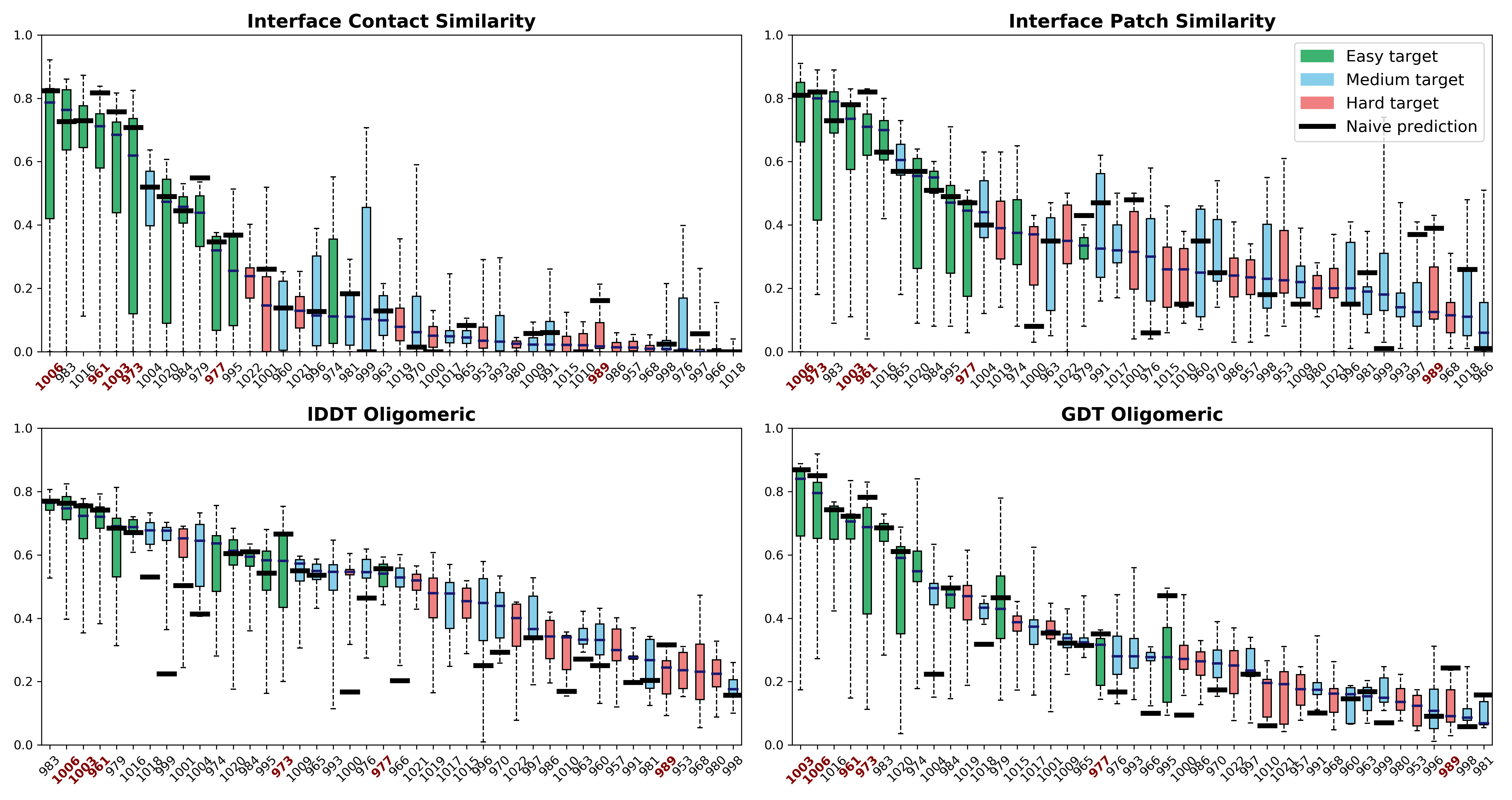}
\caption{Per-target score distributions and comparison to the baseline (na\"ive) values, if present. The targets for which the median prediction is worse than the baseline in each score are labeled in red.}
\label{figure:2:performance}
\end{figure}

We present detailed score distributions for all targets in Figure \ref{figure:2:performance}, each panel corresponding to one of the 4 scores used. We used the \texttt{Seok-naive\_assembly} method \cite{lensink2018challenge} as an indication of baseline for each target. In order to qualitatively analyze the predictions outcome, we consider a target to be solved if there exist models for which all four scores (ICS, IPS, lDDTo, GDTo) have values greater than 0.5. It follows that 9 assembly targets out of 42 are solved in CASP13: \textbf{T0961o}, \textbf{T0973o}, \textbf{H0974}, \textbf{T0983o}, \textbf{T1003o}, \textbf{T1004o}, \textbf{T1006o}, \textbf{T1016o}, \textbf{T1020o} (Figure \ref{figure:2:performance}). However, 4 of these are also solved by the baseline method. \textbf{T1004o} is a notable improvement on the baseline, as it had two partial assembly templates (PDB IDs 5EFV and 5M9F), which most groups successfully combined. In contrast to the results of tertiary structure prediction in this round of CASP, absence of detectable assembly templates with near-complete coverage guarantees absence of good models.

\begin{figure}
\centering
\includegraphics[width=0.7\textwidth]{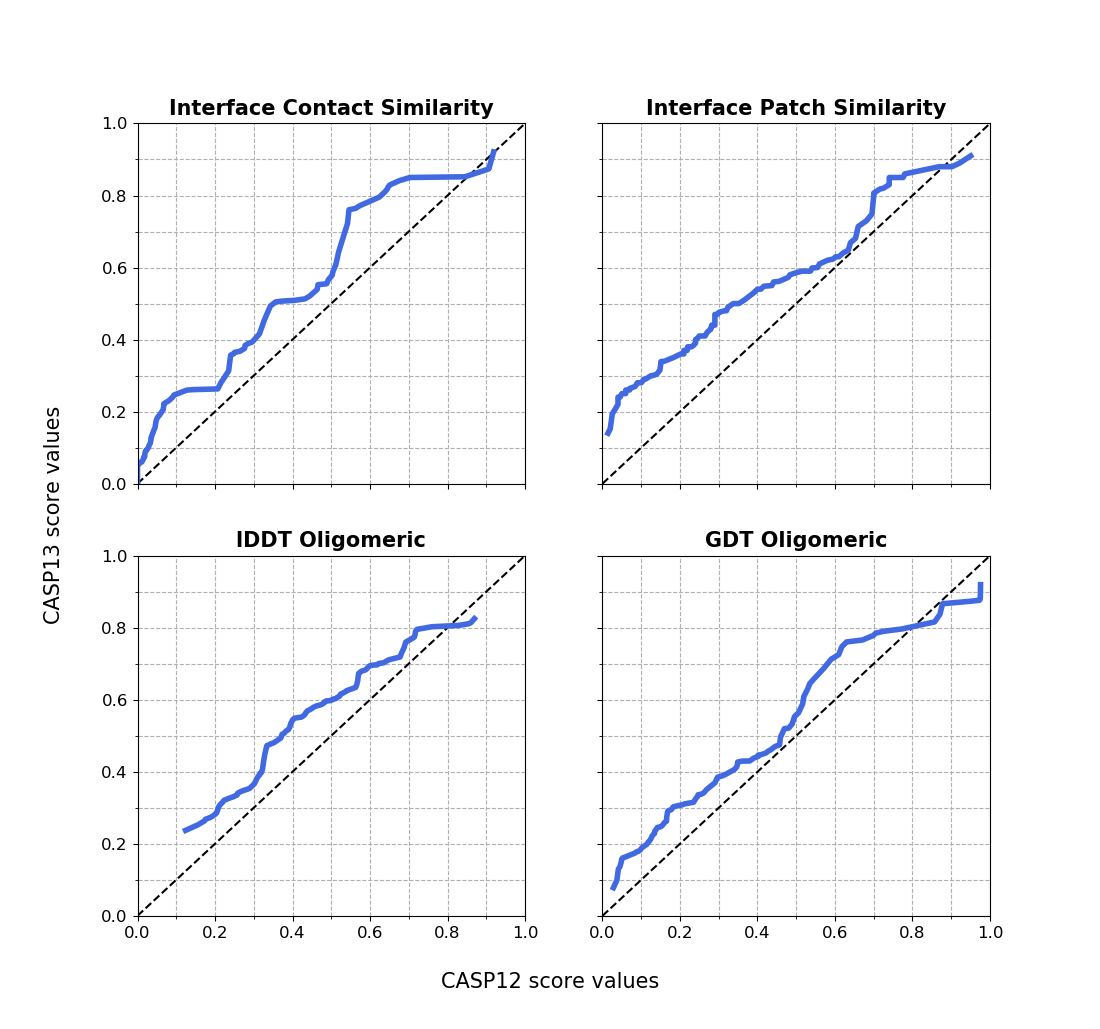}
\caption{Performance comparison between CASP13 and CASP12. 5 top predictions per target (maximum 1 per group) were selected for each score from CASP12 and CASP13 submissions. The scores were matched by percentiles and plotted as CASP12 ($x$ axis) \emph{vs.} CASP13 ($y$ axis). Values above the diagonal correspond to improvement in CASP13.}
\label{figure:4:CASP12vsCASP13}
\end{figure}

Using the same criteria as above, we find that 6 (easy) targets out of 30 were solved in CASP12 -- the same proportion as in CASP13. To evaluate the progress quantitatively, we assume that the difficulty of the assembly targets in CASP12 and CASP13 has roughly the same distribution (evidence in [ref this year's domain prediction assessment]), and compare the relative performance of the predictors by matching score percentiles. For example, GDTo value of 0.5 in CASP12 is at the 76\textsuperscript{th} percentile of all best predictions. In CASP13, the 76\textsuperscript{th} percentile corresponds to the GDTo value of 0.55, which indicates 5\% improvement. Figure \ref{figure:4:CASP12vsCASP13} reveals the complete picture of such analysis and shows 5-15\% improvement for all scores across the board. 

Finally, the CASP13 group ranking is shown in Figure \ref{figure:3:rankings}. The \texttt{Venclovas} group consistently outperformed the rest in all difficulty classes, followed by \texttt{Seok} and \texttt{BAKER}. Success of the top-performing groups appears to be in large part due to the human intervention, as all participating servers are ranked similarly to the na\"ive strategy.

\subsection{Prediction highlights}

\begin{figure}
\centering
\includegraphics[width=0.9\textwidth]{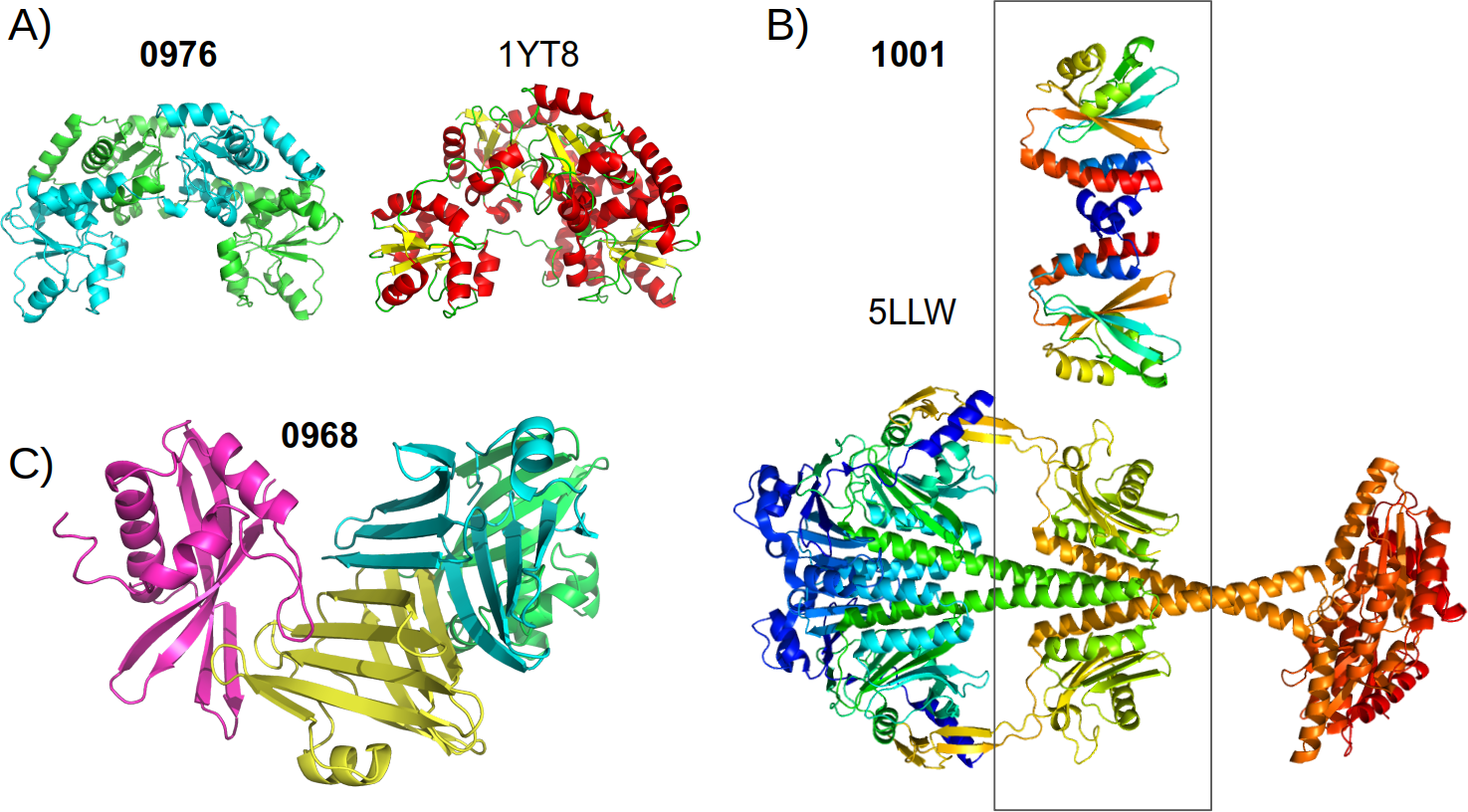}
\caption{Prediction highlights. A) The homodimeric target \textbf{T0976} and the monomeric template that matches the global arrangement of the 4 domains, B) Homodimeric target \textbf{T1001} and the template PDB entry 5LLW, a much larger protein, the highlighted central domain has a very close tertiary structure and a similar interface region. C) The A2B2 heterotetramer \textbf{T0968} with a main homomeric interface (cyan and yellow chains) via beta pairing, composing a large beta sandwich. The other subunit attaches on either side of the beta sheets.}
\label{figure:7:highlights}
\end{figure}

An interesting and quite successful prediction target was \textbf{T0976}. The homodimer target is composed of 4 copies of a well known domain with many templates available in the PDB (CATH superfamily 3.40.250.10, Oxidized Rhodanese domain 1 \cite{dawson2016cath}). However, there were no templates with this particular dimer. Rather, a monomeric template (PDB ID: 1YT8) had a similar overall arrangement of the 4 domains with interdomain interfaces resembling the dimeric interface in the target (see Figure \ref{figure:7:highlights}A). Groups like \texttt{D-Haven}, \texttt{ZouTeam} and \texttt{ClusPro} achieved relatively good scores for the dimeric interface and for the assembly.

Target \textbf{T1001}, classified as difficult, was another success story from predictors. A good dimeric template exists in the PDB (PDB ID: 5LLW), however, the matching domain in 5LLW is only a small part of the full length protein (Figure \ref{figure:7:highlights}B) and importantly contains a very long insertion when compared to \textbf{T1001}. Indeed, HHpred is not able to find either this or a tertiary-only template (PDB ID: 3OOV) when submitting different subsets of the target sequence. Relatively good predictions were submitted by \texttt{Seok} and \texttt{BAKER} groups. 

An example of an unsuccessful multimeric prediction was \textbf{H0968}, classified as difficult due to lack of assembly templates and with both monomers being FM targets. The subunits were well modelled by a few groups, presumably aided by contact prediction. However there was essentially no group that came close to either of the two interfaces present in the target (Figure \ref{figure:7:highlights}C). Nevertheless, some groups could predict interface contacts for this target's homomeric interface, as detailed in the \emph{Contact Prediction} section below.

\subsection{Importance of quaternary modelling}

\begin{figure}
\centering
\includegraphics[width=0.7\textwidth]{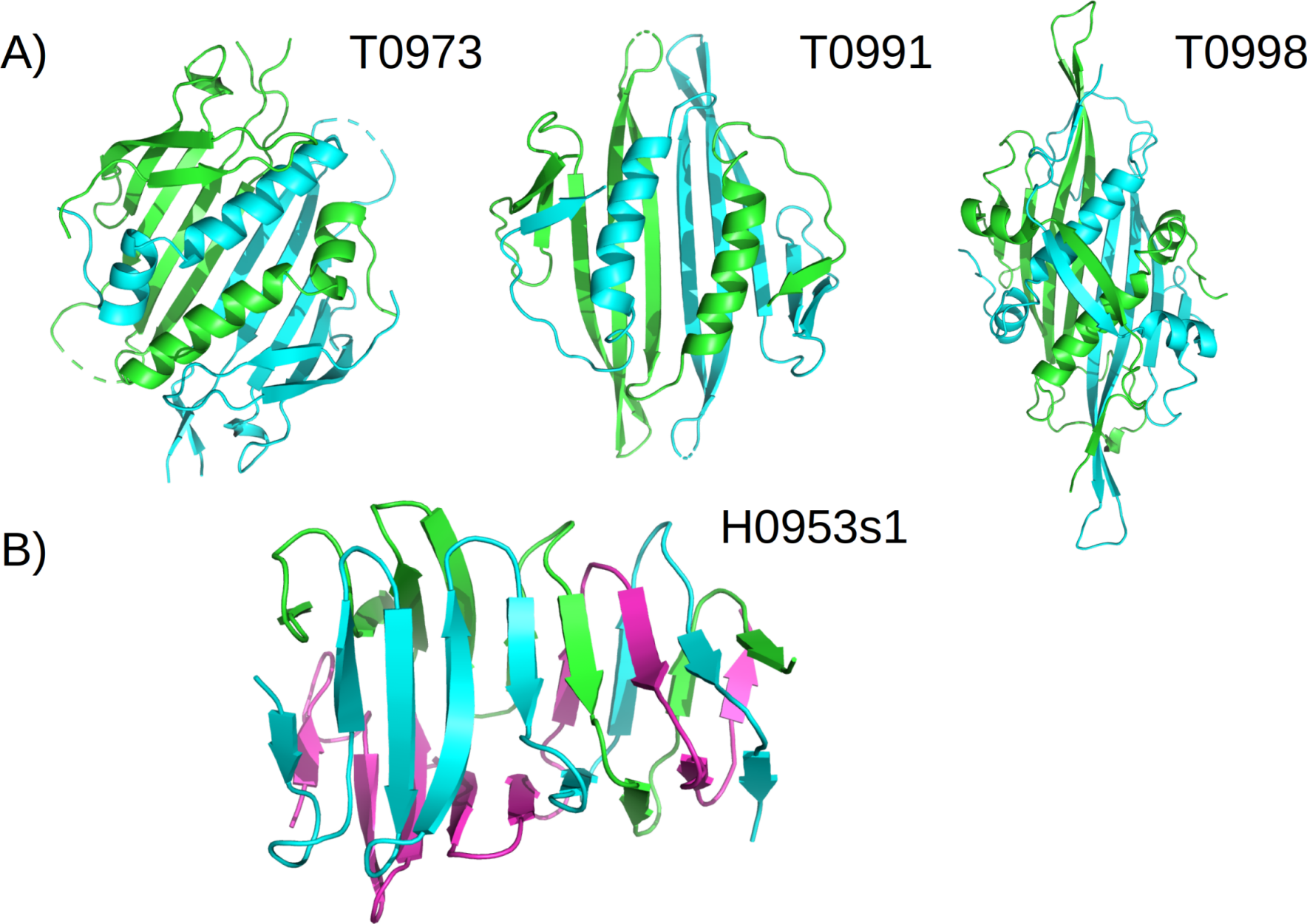}
\caption{Importance of quaternary modelling. A) Targets \textbf{T0973}, \textbf{T0991} and \textbf{T0998} with very large dimeric interfaces and the main hydrophobic core split at the interface. The best regular prediction GDT\_TS scores for their corresponding monomeric evaluation units were: 82.62 for T0973-D1, 37.16 for T0991-D1 and 35.54 for T0998-D1. B) Trimeric part of target \textbf{H0953} showing the intertwined beta-strand geometry in the C-terminal half of the fold.}
\label{figure:5:quaternary}
\end{figure}

While analyzing the results, we noticed a tendency in how the quaternary structure is handled by the predictors, in particular those who did not participate in the assembly category. Most groups seemingly split the problem into two consecutive steps: 1) modelling the subunits, 2) modelling the complex. However, results from this CASP show that such strategy is flawed. This can be appreciated very clearly in multiple targets (Figure \ref{figure:5:quaternary}) which we discuss below.

\textbf{T0973}, \textbf{T0991} and \textbf{T0998}: all 3 targets have similar folds and dimeric quaternary structures. The dimeric interface is formed by the swapping of a helix folding onto the beta sheet of the other monomer, with an enormous buried surface area resulting in an intimate and very stable dimer\footnote{Indeed, quoting Kaspars Tars (Latvian Biomedical Research and Study Centre) who provided the experimental structure: \emph{"Monomers do not exist in a free state, so modelling a monomer structure makes no sense. (...) The hydrophobic core of the protein is in part composed of inter-monomer contacts in dimer."}}. However, the evaluation unit for the regular prediction was the full monomer (including the swapped helix) in all 3 cases. Unsurprisingly, these targets received poor overall predictions. A good quaternary template was available for the target \textbf{T0973}, which resulted in some modellers achieving good scores. Notably, the best performing group in the regular category, \texttt{AlphaFold}, did not use templates explicitly and showed poor performance for \textbf{T0973} (GDT\_TS=32.62).

Target \textbf{H0953} is an A3B1 multimer, composed of a trimeric part with a beta helix fold attached to a monomeric receptor recognition protein. The trimer consists of single-chain beta sheets in the N-terminal and of interdigitated beta strands coming from each of the chains in the C-terminal. The interface buried area is not exceptionally large but the intertwining geometry makes it an obligate multimer. Again in this case, the evaluation unit (\textbf{T0953s1-D1}) was assigned to a single full-length monomer out of the trimer. This resulted in overall bad predictions in the C-terminal region for regular category models. \texttt{BAKER} is the only group that comes close to a reasonable prediction for the C-terminal.

Other examples are \textbf{T0981}, \textbf{T0989} and \textbf{H0957}. Without going into detail, all of these had relatively low-quality predictions due to treating the chains as completely independent folding units. 

\subsection{Contact predictions for homomeric interfaces}

\begin{figure}
\centering
\includegraphics[width=0.9\textwidth]{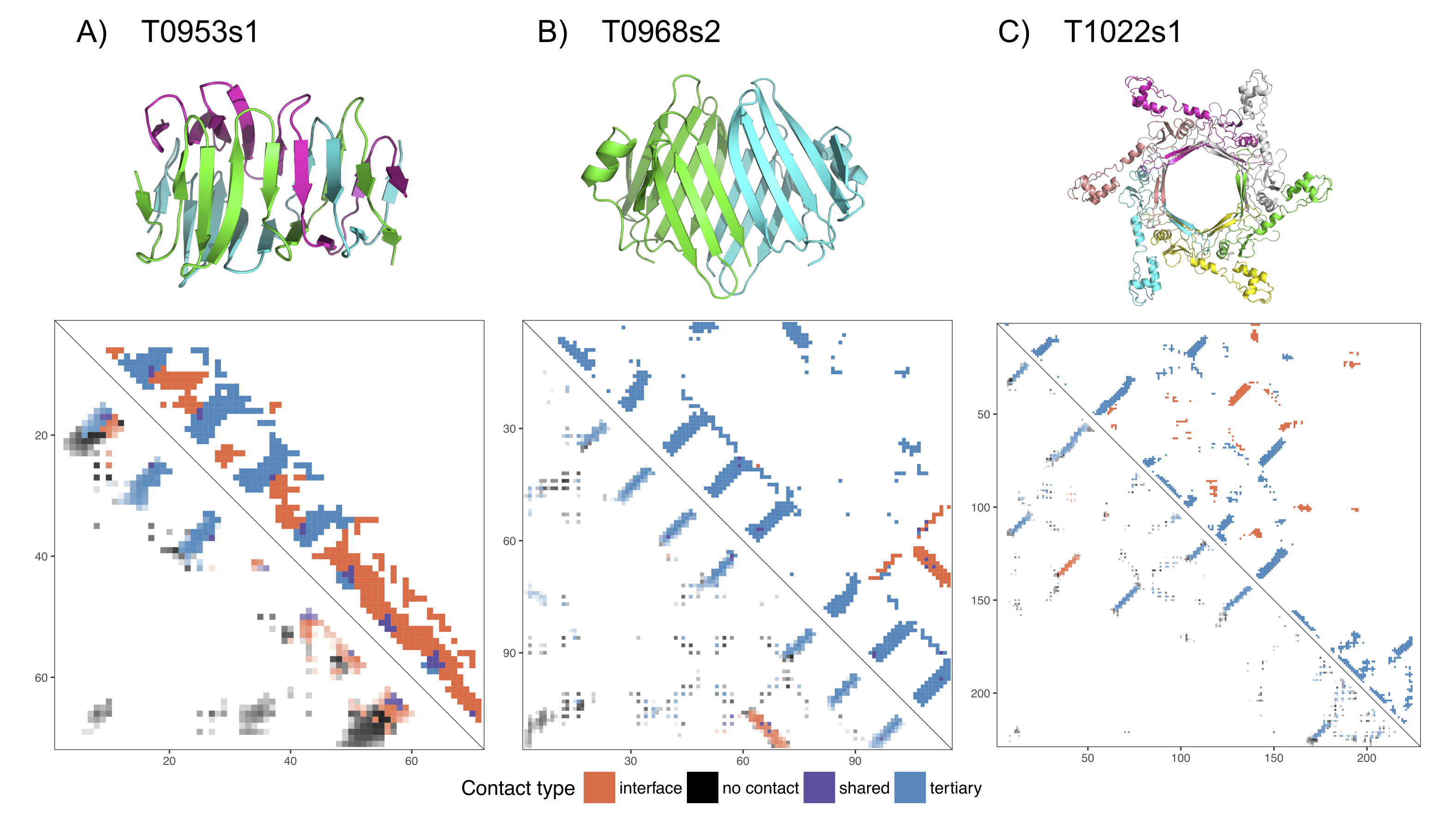}
\caption{Homomeric interface contacts (upper-right of the contact matrix) and best interface contact predictions (lower-left) for three CASP13 FM targets: A) interdigitated trimer \textbf{T0953s1} and prediction by group RR106; B) dimeric interface (isologous) of \textbf{T0968s2} and prediction by group RR036; and C) hexameric subunit \textbf{T1022s1} and prediction by group RR164.}
\label{figure:6:interface}
\end{figure}

Next, we looked whether contact predictions are in some way useful for quaternary structure modelling. Although interface contacts are not considered in the contact prediction category in CASP13 [ref Fiser 2019], homomeric interfaces are formed by contacts within a single target and should therefore be accounted for. In total, 37 CASP13 targets form homomeric interactions, which in average account for 13\% of all contacts in the target, ranging from 2\% to over 50\% (Figure \ref{figure:S1:contacts}). 
To our surprise, we find that homomeric contacts are usually among the top ranked predictions from the best groups in each respective target. In the examples shown in Figure \ref{figure:6:interface}, good predictions exist for both the tertiary and interface contacts. They are regarded as false positives in current evaluation. In fact, we find that considering homomeric contacts would have changed the group ranking for contact prediction of some targets, e.g. \textbf{T0968s2}. In view of these results, future CASP editions should consider evaluating homomeric contacts.

Homomeric interface contacts also present a challenge for protein structure modelling from contact matrix predictions, since currently most regular predictors try to fold a single subunit. The additional interface contacts in the matrix would impose unrealistic constraints between residues in the folding protocol, similarly to false positives, known to negatively affect 3D reconstruction \cite{duarte2010optimal, sathyapriya2009defining}. Modellers would need to disentangle intra-chain from inter-chain contacts in the matrix and adapt their pipelines to fold multiple chains according to the given stoichiometry, similar to what has been done for heteromeric interface predictions \cite{hopf2014sequence, ovchinnikov2014robust}.

Among all types of homomeric interactions, isologous interfaces (as found in cyclic dimers and dihedral symmetries) present yet another challenge for protein assembly modelling from contact predictions. Due to the 2-fold symmetry, many of the contacts at the interface, specially those close to the axis of symmetry, will be between the same residues (residue interacting with itself in another subunit) or residues very close in sequence, which are excluded by design from contact predictions. For example, this is the case for the homodimeric interface in target \textbf{T0968s2}.

\subsection{Data-assisted predictions and assemblies}

A total of 7 assembly targets were also released as ‘data assisted’ targets (Fig. \ref{figure:S2:assisted}), a category that attempts to evaluate advances in integrative modelling methods \cite{ogorzalek2018small}. SAXS data was collected for all 7 of the targets, whilst cross-link data was collected for 5 of them and NMR data for 1 (\textbf{H0980}). The experimental details and data-assisted specific assessment is discussed in the respective papers [refs Hura 2019, Fiser 2019, Montelione 2019]. Here, as part of our assembly analysis, we looked into how the data-assisted assembly predictions compare with the regular ones, using the regular evaluation strategy.  All 7 targets were selected from the difficult group, for which there is little homology information available to perform traditional modelling. SAXS data has the potential to provide valuable information about the global shape of the assemblies and thus should be particularly helpful for this category. At the same time, cross-linking and NMR data can  provide information on the inter-chain interfaces, potentially helping the assembly modelling process. 

Figure \ref{figure:S3:assistedscores} presents the evaluation of all the targets on the 4 scores used here (see Methods). The score ranges for all of them are not significantly different from the regular predictions.
Barring target \textbf{X0957} (Fig. \ref{figure:S4:xlinkexample}), no systematic improvement is detectable in this experiment. The reasons appear to be twofold. First, difficulty of the targets may have limited the search space of the prediction methods too early in the pipeline (Fig. \ref{figure:S5:successdiversity}). Second, the groups with the best non-assisted predictions generally did not participate in the data-assisted category, which limits comparability of the outcomes between the categories.

\section{Conclusions and Outlook}

We have presented the CASP13 assembly category assessment, the second edition of CASP with a dedicated assembly category. We have seen significant increase in participation, indicating more interest in quaternary structure modelling, a trend that can only be beneficial to the further development of methods. In addition, quality of the predictions consistently increased as well. We are hoping that the trend will continue in the next CASPs and that quaternary structure modelling becomes mainstream. Unfortunately, predictions in the regular categories are still not taking into account quaternary structure as an essential part of their modelling pipelines. We also showed that contact prediction for homomeric interfaces is already surprisingly successful, an aspect likely ignored by both predictors and assessors at the moment.

We still see room for improvement in several places. Automation is rather limited in this category. For instance, only 2 servers (Swiss-Model\cite{waterhouse2018swiss} and Robetta \cite{kim2004protein}) participate in the multimeric section of the fully automated CAMEO experiment \cite{haas2018continuous}. The sophistication of the methods in assembly modelling is falling behind traditional tertiary modelling. Specifically, we have not seen much utilization of the machine learning methods, popular in the tertiary structure and contact prediction categories. It appears that traditional homology modelling still dominates the field.

In conclusion, we would like to emphasize that quaternary modelling is intrinsic to the protein modelling problem and must be considered from the outset in the design of modelling pipelines. Correspondingly, a CASP evaluation unit should match the functional form of a protein structure, be it a monomer or an assembly, with consistent metrics throughout.

\section*{Acknowledgements}
We would like to thank the CASP management committee for organization and support. We are grateful to Chaok Seok for contributing the na\"ive prediction method. We thank Susan Tsutakawa, Gregory Hura, Gaetano Montelione and Andras Fiser for their help in interpreting data-assisted predictions. We thank Spencer Bliven for discussions of the CASP12 assembly prediction results.

\bibliography{sample}
\end{multicols}

\pagebreak

\pagebreak
\renewcommand{\thefigure}{S\arabic{figure}}
\setcounter{figure}{0}

\begin{figure}
\centering
\includegraphics[width=0.6\textwidth]{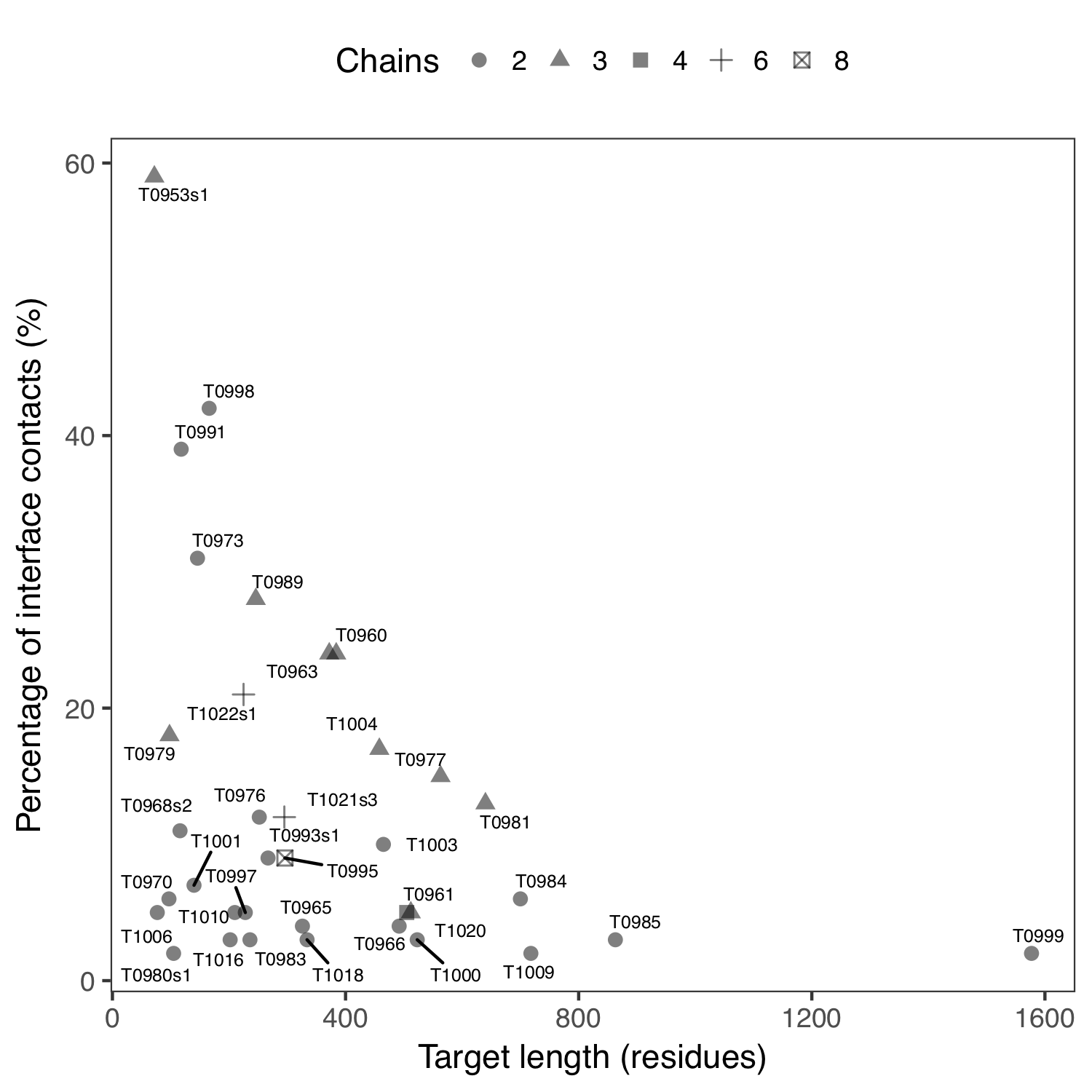}
\caption{Percentage of homomeric interface contacts in CASP13 targets.}
\label{figure:S1:contacts}
\end{figure}

\begin{figure}
\centering
\includegraphics[width=0.9\textwidth]{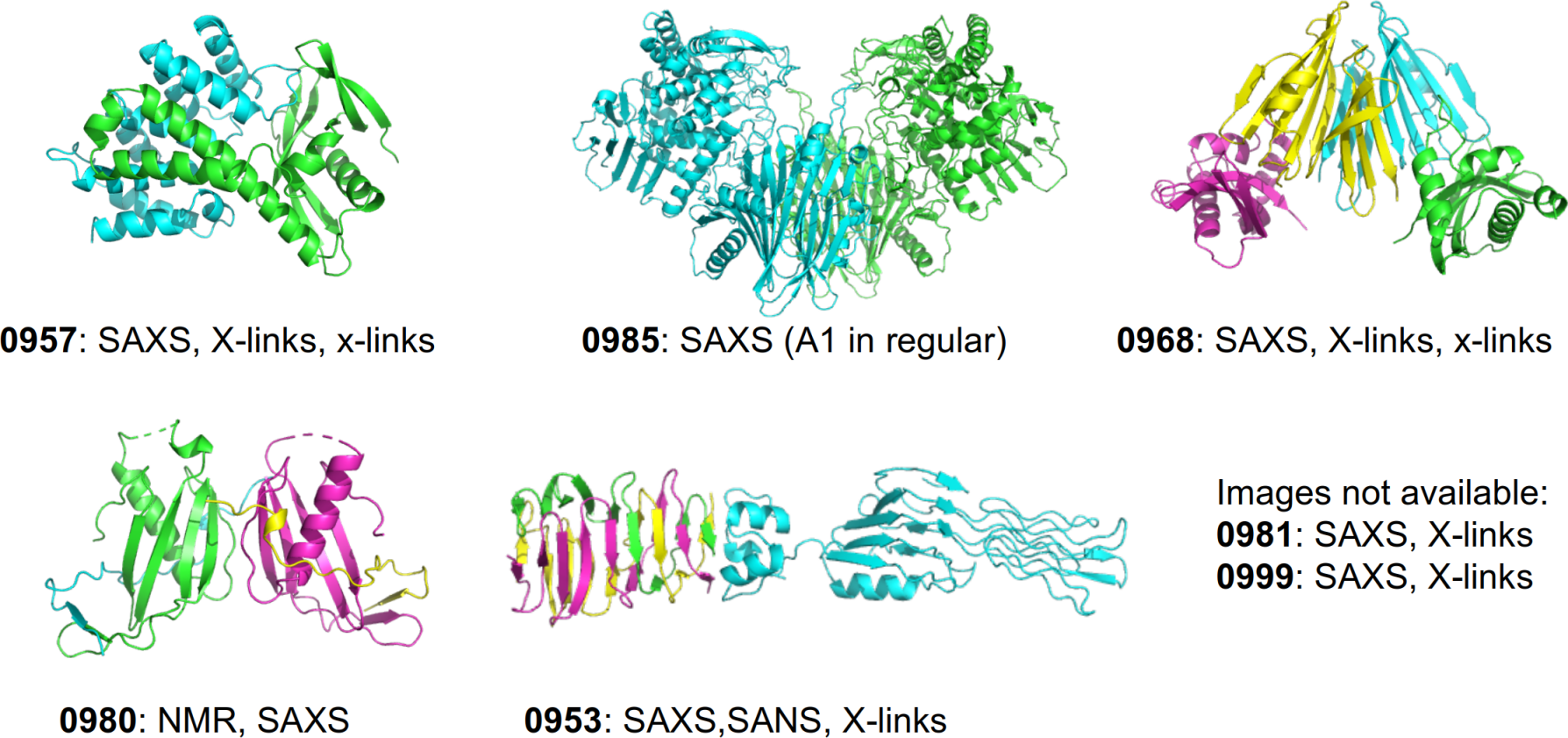}
\caption{Data-assisted targets.}
\label{figure:S2:assisted}
\end{figure}

\begin{figure}
\centering
\includegraphics[width=1\textwidth]{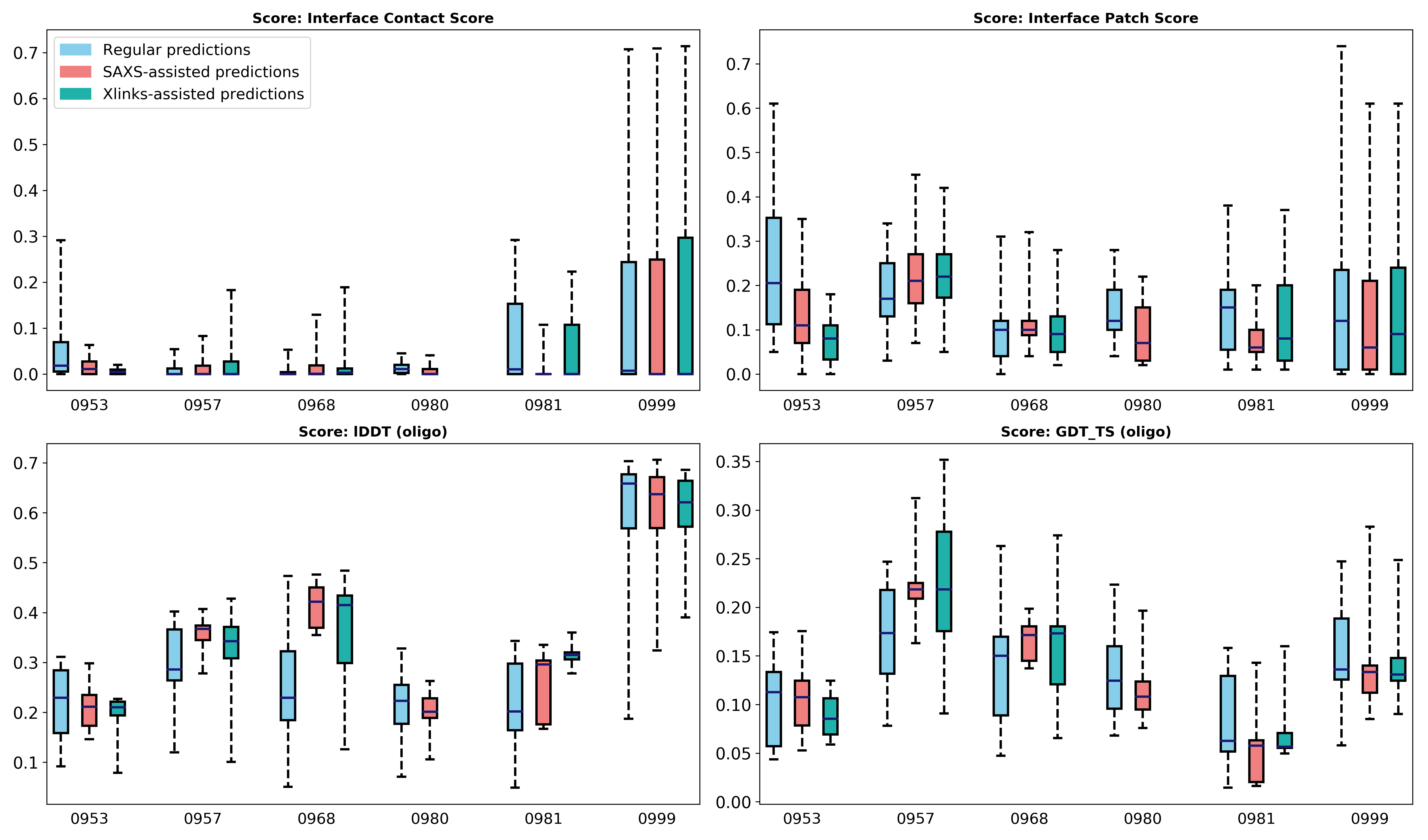}
\caption{Score distributions for all predictions of data-assisted and the corresponding non-assisted targets. Two types of crosslinks and two types of scattering datasets are merged for the purpose of this figure.}
\label{figure:S3:assistedscores}
\end{figure}

\begin{figure}
\centering
\includegraphics[width=0.6\textwidth]{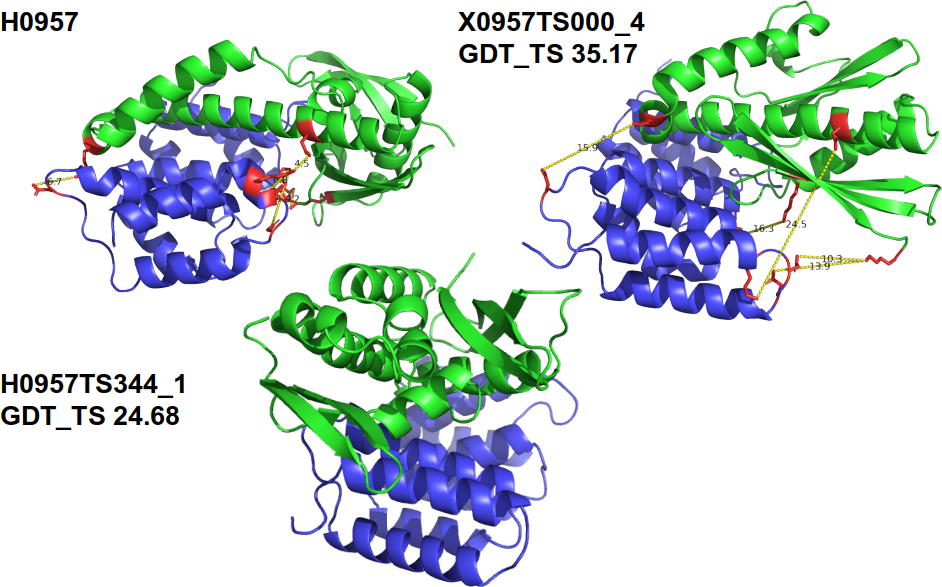}
\caption{Target X0957 shows improvement across all scores considered due to several fortunate intermolecular crosslinks. Crosslinked residues in the target and the assisted prediction are highlighted in red and connected with a dashed yellow line. Crosslinks between missing residues are not shown. Best regular prediction (bottom) has a significantly lower GDT.}
\label{figure:S4:xlinkexample}
\end{figure}

\begin{figure}
\centering
\includegraphics[width=0.9\textwidth]{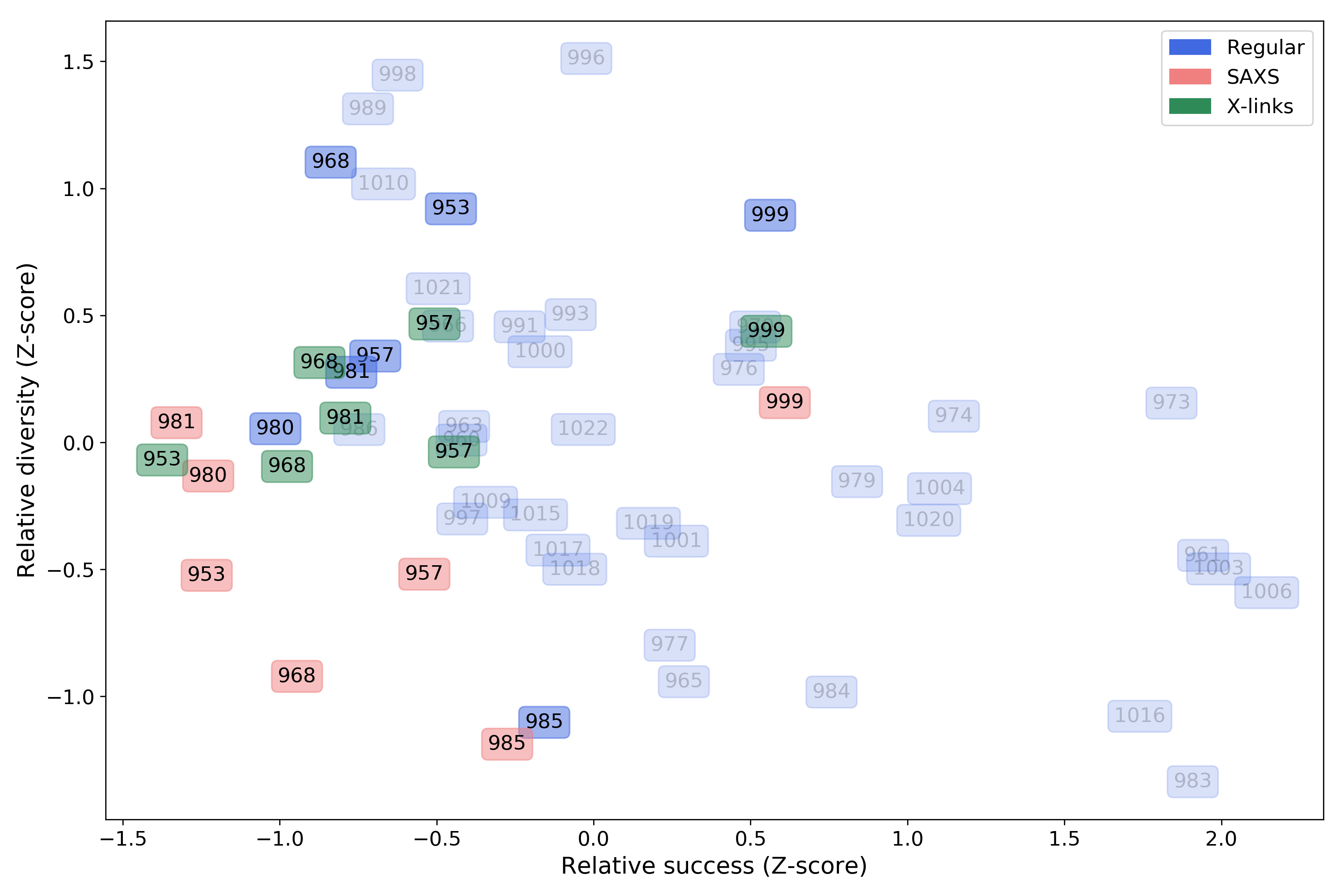}
\caption{Relative scores vs relative score diversity, by target. 95th percentile of each score was calculated for each target and taken to represent a 'good' prediction for this target. Sum of the corresponding $Z$-scores is shown on axis $X$ and represents relative predictions success for the targets. Coefficient of variation (standard deviation divided by mean) of the scores per target, normalized to $Z$-scores, is shown on axis $Y$ and represents diversity of the predictions.}
\label{figure:S5:successdiversity}
\end{figure}

\end{document}